\documentclass[12pt]{article}
\usepackage{latexsym,epsfig}
\usepackage{titlesec}
\usepackage{hyperref}
\usepackage{color}

\begin{document}
\centerline{\bf \large Do We Live in an Antigravity Universe?}
%
%

\bigskip

\centerline{S. Menary}
\smallskip

\centerline{Dept. of Physics \& Astronomy}
\centerline{York University, 4700 Keele Street}
\centerline{Toronto, ON M3J 1P3, Canada}
\bigskip

\date{\today}
\begin{abstract}
In a previous article\cite{antiart} I showed that, in the context of antigravity (i.e., matter and antimatter repel gravitationally), quark/lepton mass-energy is matter and antiquark/antilepton mass-energy is antimatter while the mass-energy of the intermediate vector bosons (e.g., the photon and $W^+$) also has to be considered matter. One consequence of this is that the antiproton (and, hence, the antineutron) is dominantly composed of matter since some two-thirds of its mass is gluonic. Under this premise I found that the gravitational acceleration of antihydrogen would be $a_{\bar{H}}=(0.33^{+0.23}_{-0.11})g$. This is to be compared to the ALPHA-g result\cite{ALPHA-g} of $a_{\bar{H}}=(0.75\pm 0.13~({\rm stat.+syst.})\pm 0.16~({\rm simulation}))g$.
In this article I explore the cosmological implications of this definition of matter and antimatter.
This leads to a quite different antigravity universe than previous analyses with, for example, equal amounts of hydrogen and antihydrogen but far fewer antistars than stars. I examine the observations used to extract various parameters of the $\Lambda$CDM model of the universe and show that they are potentially consistent with the characteristics of the antigravity universe. A precise version of antigravity, one which must include General Relativity, is needed to generate a fully consistent, predictive model of the antigravity universe but even without it the antigravity scenario outlined here naturally leads to a rich phenomenology including different acceleration and expansion rates between the early universe and the present, MOND-like galaxy rotation curves, cosmic voids, and more.
\end{abstract}
\clearpage
\hrule

\tableofcontents
\vskip 1cm

\hrule
\bigskip

Antigravity is the proposition that matter and antimatter gravitationally repel each other\footnote{See \cite{antiart} for a brief summary of the history and present status of theories of antigravity.}. It is clear that to investigate the implications of antigravity requires a precise understanding of what exactly one means by the terms "matter" and "antimatter". In \cite{antiart} I showed that 
the mass-energy of the fundamental fermions of the Standard Model of Particle Physics (i.e., quarks and leptons) and the intermediate vector bosons (the photon, $W^\pm$, $Z^0$, and the 8 coloured gluons) act gravitationally as matter while the mass-energy of the antiquarks and antileptons (the fundamental antifermions) constitute antimatter. 

Previous antigravity analyses (see, for example, \cite{dim}\cite{vdark}\cite{haj}\cite{char}) assumed that equal amounts of matter and antimatter were created in the Big Bang. What is really meant is that equal numbers of fermions and antifermions were created but, with the precise definition of what constitutes matter and antimatter given here, even just considering photons (which count as matter) we have at the instant of the Big Bang more matter than antimatter in the universe\footnote{Yet, to repeat, and this can't be emphasized enough, there are equal numbers of fundamental fermions and antifermions.}. This conclusion is independent of any particular theory of antigravity and must be taken into consideration in any analysis involving antigravity. 

Furthermore, if antigravity is active then it is no longer possible to consider composite systems (i.e., those containing both matter and antimatter) in as straightforward a manner as in standard gravity. An analogy would be an ionized atom in an electric field. The ion contains positive and negative electric charges but it is the net electric charge which is important when considering the dynamics of the system. Of particular relevance for the antigravity universe is the antiproton (and antineutron) which, because the gluonic mass-energy component acts gravitationally as matter\cite{antiart}, is roughly two-thirds matter and one-third antimatter. This has profound consequences for cosmic evolution and structure.
\section{Aspects of the Antigravity Universe}

Everything I've written up to this point is independent of how antigravity is realized. To make specific predictions does depend on a particular mechanism for activating antigravity. It is clear, however, that if antigravity is acting then the standard cosmological model, $\Lambda$CDM, is at best incomplete and certainly does not contain the terms necessary to describe the effects of antigravity on cosmological structure and dynamics. For example, the basic assumptions upon which the Friedmann-Lema\^itre-Robertson-Walker metric is derived are not true in the antigravity universe since, for example, it is not homogeneous.

\subsection{Matter-Antimatter Asymmetry and Hadronization}

As stated above, we begin with the matter-antimatter asymmetry being there right from the instant of the Big Bang. There are indeed equal amounts of fundamental fermions (matter) and antifermions (antimatter) created at the Big Bang but also some number of intermediate vector bosons (or the $X$ and $Y$ bosons from GUT theories) which act gravitationally as matter. Matter and antimatter would presumably congregate into separate regions. The repulsion of regions of matter and antimatter has two nice features; 1) it leads naturally to an accelerated expansion of the universe (i.e., acts like Dark Energy) and 2) an equal number of fundamental fermions and antifermions is maintained and since they are spatially separated, you don't require some mechanism (e.g., the Sakharov conditions) to eliminate the antimatter left over after the annihilations that lead to the Cosmic Microwave Background. 

A curious twist in this scenario occurs at the time of hadronization (approximately a microsecond after the Big Bang in the standard cosmology). The nucleons form as usual but the antimatter dominated regions go through a kind of phase change. That is, the antifermions are indeed pure antimatter but the antiquarks coalesce into antinucleons which are roughly two-thirds matter. So now even the previous pockets of antimatter become dominantly matter and the universe is matter dominated since the baryons are completely matter and the equivalent number of antibaryons are now dominantly matter.
The antiprotons would combine with positrons to make antihydrogen and there would also presumably be some antihelium production\footnote{The STAR Collaboration at RHIC measured the strength of the interaction between antiprotons\cite{STAR}(2015) and found it to be consistent with that between protons. And so, just as $CPT$ Invariance demands, the antihelium nucleus is taken to have the exact same properties as helium.}. 

\subsection{Where Are All the Antistars?}

The fact that antinucleons\footnote{I will continue to say antinucleon and nucleon although by this point in the evolution of the universe they are dominantly bound states of antihydrogen and hydrogen with some antihelium and helium thrown in.} are approximately two-thirds matter and only one-third antimatter\footnote{Specifically I showed in \cite{antiart} that the antimatter fraction of the antiproton mass is $\bar{f}_{\bar{p}}=0.33^{+0.06}_{-0.10}$. This fraction should be very close to the same for the neutron.} means that the strength of the gravitational interaction between them is weaker than that between nucleons and, hence, antistar formation does not proceed as it does for stars and there are invariably less antistars than stars. This is a general conclusion again independent of any particular antigravity model. 

To make specific predictions I will draw on the work of Villata\cite{vill1} (I summarized it in \cite{antiart}), in which he found that matter-antimatter gravitational repulsion is already a feature of the General Theory of Relativity if $CPT$ is conserved. He also showed that photons (I expanded this to include all intermediate vector bosons) act gravitationally as matter\cite{vill4}, consistent with what is observed (e.g., bending of light by the Sun). Based on this I derived the general force equation between two states\cite{antiart}. To recap that analysis, representing the mass $M$ of a composite state by $M(f+\bar{f})$ where $f$ and $\bar{f}$ are the matter and antimatter fractions, respectively, it followed that in the Newtonian limit the gravitational force between two states is:
\begin{equation}
    F_{12}=\frac{-GM_1M_2\left(1-2(\bar{f}_1+\bar{f}_2)+4\bar{f}_1\bar{f}_2\right)}{r_{12}^2}
\end{equation}
Note that, as must be the case, for pure matter ($\bar{f}=0$) and antimatter ($\bar{f}=1$) states you have:
$$F_{ff}=F_{\bar{f}\bar{f}}=-F_{f\bar{f}}=-G\frac{M_1M_2}{r_{12}^2}$$
Substituting $\bar{f}$ = 1/3 for an antinucleon and $\bar{f}$ = 0 for the nucleon results in:
\begin{equation}
F_{\bar{n}\bar{n}}=\left(\frac{1}{3}\right)F_{\bar{n}n}=\left(\frac{1}{9}\right)F_{nn}
\end{equation}
where I've taken $M_{\bar{n}}=M_n$.

A gravitational force between antinucleons that is one-ninth that between nucleons translates into a formation time at least three times greater for antistars than stars\footnote{It's a fairly straightforward calculation to show that the time for an antinucleon to move under the force of gravity a specified distance closer to another antinucleon, call this time $t_{\bar{n}\bar{n}}$, is related to the time for the same scenario involving two nucleons, $t_{nn}$, in the following way: $t_{\bar{n}\bar{n}}/t_{nn}$ = $\sqrt{F_{nn}/F_{\bar{n}\bar{n}}}$ = $\sqrt{9}$ = 3.}. The situation is further complicated by the fact that the nucleon-antinucleon gravitational force is three times that of the antinucleon-antinucleon force and so antistar formation depends also on the spatial distribution not only of antinucleons but also nucleons. Finally, at what rate would antistar production proceed when we have the case that the radiation pressure due to nuclear fusion processes is the same in antistars as stars but the gravitational force for antistars is one-ninth that for stars? It is impossible to precisely predict what the antistar production rate will be without a simulation which takes into account the distributions of nucleons and antinucleons and the relative forces involved but it is clear that,  
even though there are as many antinucleons as nucleons in the antigravity universe, there are potentially far, far fewer antistars than stars. And the weaker gravitational attraction of antinucleons to each other versus with nucleons could lead to, for example, galaxies being immersed in a diffuse cloud of antinucleons. 

Note that, since there is negligible antistar production, all the conclusions from Big Bang Nucleosynthesis (BBN) should remain valid. That is, you would expect the production of heavy elements to proceed just as described in BBN but there is not a corresponding production of heavy anti-elements as there are few antistars. In other words, there won't be any antiplanets ... or antipeople! 


\subsubsection{Cosmological Anti-Antigravity Arguments Revisited}

There were a number of papers ruling out an antigravity universe on cosmological grounds\cite{steig1}\cite{rujula}\cite{steig2}. Central to their argument is the absence of gamma rays from antistar-star annihilations (see also \cite{gao}). Many tried to salvage what was often called the "Baryon Symmetric Big Bang" by postulating that the antistars/antiplanets/antigalaxies existed in spatially separated pockets\cite{gao}\cite{blinn}\cite{duda}. The fundamental assumptions in all of these is that there are equal numbers of antistars as stars.
However, as we've seen, equal numbers of antinucleons and nucleons does not lead to equal numbers of antistars and stars and, in fact, the antistar population could well be negligible compared to that of the stars. Further, whatever antistars there are contain more matter than antimatter (since antihydrogen is more matter than antimatter), and photons act gravitationally as matter so, even if there were an equal number of antistars as stars, the universe would still contain more matter than antimatter.

\subsubsection{Vestiges of Antimatter in our Galaxy}

In the antigravity universe, galaxies repel positrons and attract antinucleons (and so antihydrogen and antihelium) albeit with a force a third of that to nucleons so one would expect some small amount of antimatter in a galaxy. 

Evidence of a surfeit of positrons emanating from the galactic centre comes from the observation of the 511 KeV photons from electron-positron annihilation\cite{jean}. In \cite{vincent}(2015) they found, "$10^{43}$ positrons per second annihilate in a compact spherical region around the centre of the Milky Way. At present, known astrophysical sources cannot account for this signal." In the antigravity universe Black Holes can act as positron sources although it is not a given (see Appendix A for a discussion of this). The AMS experiment also finds excesses of positrons, antiprotons, and antihelium\cite{Ting}(2018) above expected levels from cosmic rays. 
There are quite a number of papers trying to explain these excesses using known processes while there is also recognition that they may be harbingers of new physics or signals for dark matter\cite{poulin}(2019)\cite{cran}(2020). Such observations are quite natural in the antigravity universe where you would expect galaxies to be immersed in (or perhaps mostly surrounded by) a diffuse cloud of antihydrogen with some antihelium added in. 

There was also the claim of the observation\cite{antistar}(2021) of some fourteen antistars in our galaxy. The signature was gamma-ray spectra consistent with that coming from nucleon--antinucleon annihilation (see Appendix B for an extended discussion of antinucleon-nucleon annihilation). They interpreted their signal as being due to interstellar matter accreting onto an antistar whereas in the antigravity universe the opposite is just as likely, that the signal is due to interstellar antihydrogen accreting onto a normal star. The exact same spectra would result and one would expect the latter scenario to more probably occur in the galactic halo. 
In any case, it still represents evidence for some amount of antinucleons in our galaxy, either as free antihydrogen or antistars, as expected in the antigravity universe. 

\subsection{Antineutrinos -- Pure Antimatter}

The only truly free antimatter objects are antineutrinos which would be repelled by everything else in the universe other than remnant positrons which didn't combine with antiprotons to form antihydrogen atoms. Relic antineutrinos would presumably congregate into isolated pockets or form strands or streams woven into the universe. 

Black Holes would be prodigious sources of antineutrinos\footnote{See Appendix A for a fuller discussion of Black Holes.} since the neutrinos in the $\nu\bar{\nu}$ pairs produced in Hawking radiation\cite{HR} would be absorbed into the matter dominated Black Hole while the antineutrinos would be repelled. Antineutrinos from a Black Hole located at the centre of a galaxy would stream away from the galaxy\footnote{Black Holes as antineutrino sources was discussed in \cite{drag} although in this paper it is claimed that observation of these antineutrinos from the centre of the galaxy is possible evidence for antigravity while I assume that the antineutrinos are repelled by our matter-dominated Solar System and so we wouldn't observe them.} and be attracted to those pockets of the antigravity universe populated by relic antineutrinos.

To really see how these antineutrinos would be distributed again requires a proper calculation but it is clear that antineutrinos could potentially play a big role in the evolution of cosmic structure. 
   
\subsection{(Anti)Gravitational Lensing}

Regions populated by antineutrinos or by antinucleons greatly complicate the gravitational lensing interpretration of how mass is distributed in the universe. Since photons act as matter under gravitation, the force exerted on them by antinucleons is one-third that due to nucleons. Therefore the deflection angle is three times smaller (or perhaps $\sqrt{3}$ times smaller since the Einstein angle is proportional to $\sqrt{GM}$)
for lensing from antistars (or clusters of antinucleons) than from stars\footnote{Recall that the deflection angle calculated from GR is just twice that derived from Newtonian gravity\cite{soares}.}. Further, if galaxies are surrounded by, or are immersed in, a diffuse cloud of antinucleons then the lensing would be due to a configuration of stars and antihydrogen. For pockets of antineutrinos the effect is even more pronounced since photons are repelled by antineutrinos so such regions actually {\it defocus} the light coming from behind. 


\section{So, {\it Do} We Live in an Antigravity Universe?}

An alternative title for this section could be, "So, Are the Astronomical Observations Consistent with Antigravity?" 
In \cite{mod}(2012) they wrote the following about requirements of any modifications to GR;
\begin{quote}
    \begin{it}
        Of course, if we wish to account for dark energy, or solve the cosmological constant problem using modified gravity, these deviations must be manifest in
the solutions of the Friedmann equations. We must also require, however, that they do
not spoil the successful predictions of the standard cosmology, such as the abundance
of light elements, the peak positions of the CMB acoustic spectrum, or the predictions
for baryon acoustic oscillations. This requires the background FLRW cosmology of the
modified theories to closely mimic the standard evolution of $\Lambda$CDM from nucleosynthesis
through to matter domination.
    \end{it}
\end{quote}

I propose to first explore the "observational evidence" for $\Lambda$CDM and then to see if the characteristics of the antigravity universe can offer a simpler and more compact alternative explanation.


\subsection{The Standard Cosmology and Observables}
Again quoting \cite{mod}, "the assumptions that go into the standard cosmology include;
\begin{enumerate}
\item Einstein’s field equations.
\item The universality of free fall.
\item Local Lorentz invariance.
\item Three spatial dimensions (below the electro-weak scale, at least).
\item Conservation of energy-momentum.
\item Homogeneity and isotropy of space on large scales.
\item Matter fields being well modelled by fluids of dust and radiation
\end{enumerate}
{\bf These assumptions, when confronted with observations, require the presence of dark
matter and dark energy.}"

The emphasis on the last statement is mine. It is critically important to the arguments that follow to understand that the necessity to include Dark Matter and Dark Energy is underpinned by the requirement these assumptions be true. Antigravity is not in this list and so the derived parameters of the standard cosmology are not necessarily exactly those to be compared to observations. 

There are many nice descriptions of the standard $\Lambda$CDM model. For consistency I will continue to follow that given in \cite{mod} in order to examine the solutions that were derived from the above assumptions and the subsequent observables that emerge. The Robertson-Walker metric in a spherically symmetric coordinate system is
\begin{equation}
ds^2=-dt^2+a^2\left[\frac{dr^2}{1-\kappa r^2}+r^2d\Omega\right]
\end{equation}
where $\kappa$, the Gaussian curvature, is a real constant. The function $a(t)$ is the scale factor assumed to be normalized to be equal to 1 at the present moment. The Hubble parameter, $H$, is defined as
\begin{equation}
H=\frac{1}{a}\frac{da}{dt}
\end{equation}
with $H_0\equiv H(a=1)$ being the value of the Hubble parameter at the present moment. The dynamics of the scale factor is given by the Friedmann equation:
\begin{equation}
3H^2=8\pi G\sum_i\rho_i~~~\Longrightarrow~~~\rho_T\equiv\sum_i\rho_i=\frac{3H^2}{8\pi G}
\end{equation}
where $\rho_i$ are the energy densities of all the possible fluids including photons, $\rho_\gamma$, neutrinos, $\rho_\nu$, pressureless matter, $\rho_m$, and spatial curvature, $\rho_\kappa$ and $\rho_T$ is the total energy density. Often what is discussed are the relative densities
$$\Omega_i=\frac{\rho_i}{\rho_T}$$
and so the Friedmann equation can be written
$\sum_i\Omega_i=1$. Finally, if each fluid is uncoupled then energy-momentum conservation gives
\begin{equation}
\frac{d\rho_i}{dt}+3H(1+w_i)\rho_i=0
\end{equation}
where $w$ is the equation of state defined by $P=w\rho$. 

This leads to a number of parameters, used to characterize the standard cosmology, to which observations are compared. These typically include; the Hubble constant $H_0$, the baryon density $\Omega_b$, the matter density $\Omega_m$, the equation of state $w$, the fluctuation amplitude $\sigma_8$, a measure of the "clumpiness" of matter in the universe (or, as is more commonly quoted, $S_8\equiv \sigma_8\sqrt{\Omega_m/0.3}$), the optical depth to reionization $\tau$, and the scalar power law index $n_s$. 

This is a tidy picture but maybe not the whole story, as discussed in detail in \cite{abdalla}(2022) where they state;
\begin{quote}
\begin{it}
 The standard $\Lambda$ Cold Dark Matter ($\Lambda$CDM) cosmological model provides a good description of
a wide range of astrophysical and cosmological data. However, there are a few big open questions
that make the standard model look like an approximation to a more realistic scenario yet to be
found. In this paper, we list a few important goals that need to be addressed in the next decade,
taking into account the current discordances between the different cosmological probes, such as the
disagreement in the value of the Hubble constant $H_0$, the $\sigma_8-S_8$ tension, and other less statistically
significant anomalies. While these discordances can still be in part the result of systematic errors,
their persistence after several years of accurate analysis strongly hints at cracks in the standard
cosmological scenario and the necessity for new physics or generalisations beyond the standard
model.
\end{it}
\end{quote}

In the following I'll attempt to outline how the antigravity universe gives an alternative view of the observations and could perhaps alleviate the need for many of the features added to GR (e.g., Dark Energy) or currently not being well described by $\Lambda$CDM (e.g., $H_0$).

\subsection{Extracting Parameters using Observations}
 
My goal here is to try and consistently apply the characteristics of the antigravity universe to extract cosmological parameters using the astronomical observations.
One of the difficulties in comparing expectations for parameters in the antigravity universe to "observations" is that the published parameters derived from these observations required interpretation within the $\Lambda$CDM picture (see, for example, \cite{pardo}(2020)). 

Gravitational lensing analyses are more complicated in the antigravity universe than in the $\Lambda$CDM universe. As discussed in Section 1.4, the antihydrogen gravitational mass (that which is relevant for lensing) is about one-third its inertial mass and so it affects the path of EM radiation like an atom with a third the mass of hydrogen. Assuming the amount of antihydrogen is roughly the same as the amount of hydrogen then this would presumably just lead to the conclusion that the mass $M_g$ of a galaxy plus antihydrogen cloud system is about 50\% larger than lensing indicates (i.e., lensing would "sense" $M_g\approx M_H+M_{\bar{H}}/3=(4/3)M_H$). Antineutrinos act as a defocussing element and would have more effect on cosmic structure (e.g., the CMB, $S_8$, etc.). Taking this into account would inevitably change the conclusions reached in a raft of recent papers. Here are a few examples from the last couple of months alone, and in no particular order, of papers using lensing to extract information about the universe;
\begin{itemize}
    \item "Weak-Lensing Characterization of the Dark Matter in 29 Merging Clusters that Exhibit Radio Relics"\cite{Finner}
\item "Signatures of dark and baryonic structures on weakly lensed gravitational waves"\cite{Brando}
\item "Systematic effects on lensing reconstruction from a patchwork of CMB polarization maps"\cite{Ryo}
\item "Gravitational Lensing Reveals Cool Gas within 10-20 kpc around a Quiescent Galaxy"\cite{Barone} 
\item "A new non-parametric method to infer galaxy cluster masses from weak lensing"\cite{Mistele}
\item "Analytical Weak Lensing Shear Inference for Precision Cosmology"\cite{Li}
\item "Galaxy-dark matter connection from weak lensing in imaging surveys: Impact of photometric redshift errors"\cite{Navin}
\item "Peculiar dark matter halos inferred from gravitational lensing as a manifestation of modified gravity"\cite{Bilek}
\item "Kinematic Lensing with the Dark Energy Spectroscopic Instrument -- Probing structure formation at very low redshift"\cite{Xu}
\item "Weak Gravitational Lensing around Low Surface Brightness Galaxies in the DES Year 3 Data"\cite{Chicoine}
\item "Neutral hydrogen lensing simulations in the Hubble Frontier Fields"\cite{Blecher}
\end{itemize}
This is by no means a comprehensive list but illustrates the number of observatories taking data used in lensing analyses including:
\begin{itemize}
    \item The Dark Energy Survey (DES)\cite{DES}
    \item The Atacama Cosmology Telescope (ACT)\cite{ACT}
    \item The Kilo-Degree Survey (KIDS)\cite{KIDS}
    \item The Optical Gravitational Lensing Experiment (OGLE)\cite{OGLE} 
    \item The Large Synoptic Survey Telescope (LSST)\cite{LSST}
\end{itemize}

The central point (and one that I admit is extremely unsatisfying) is that it is difficult to use any parameter derived from a standard analysis of lensing data to refute (or verify) an aspect of the antigravity universe. To really generate predictions from the antigravity scenario one would have to assume distributions of antihydrogen and antineutrinos (and standard matter too, of course) and then do a lensing analysis. Just to be clear, this says nothing about the observations themselves, just the parameters extracted from those observations.
\subsection{The Antigravity Universe}
As discussed previously, the additional features of the antigravity universe which allow for a different point of view regarding a number of cosmological observables include; 
\begin{itemize}
    \item[1)] There are equal numbers of nucleons and antinucleons but, because the force between antinucleons is about one-ninth that between nucleons, there are far few antistars than stars. 
    \item[2)] The only free form of pure antimatter is antineutrinos which are repelled by everything else. 
    \item[3)] Both 1) and 2) impact gravitational lensing analyses in a non-trivial way.
\end{itemize}

\subsubsection{Cosmic Acceleration/Dark Energy and The Hubble Parameter}
The $\Lambda$CDM includes, as the name implies, Dark Energy in the form of a cosmological constant, $\Lambda$, and Dark Matter. I will discuss Dark Matter in Section 2.3.2 but here we concentrate on cosmic acceleration and the Hubble parameter. In both cases there has recently been found a certain "tension" between values for the parameters extracted from different observables. 

The different values for the Hubble constant from analyses using different methods for deriving the length scale\cite{freed}\cite{reiss1}\cite{reiss2}(2024) -- the so-called "Hubble tension" -- has become a source of much recent debate and, of course, preprints (see, as just one example of many, \cite{per}(2024))! 
As well there is some recent evidence that the accelerated expansion of the universe has not been constant (i.e., is not due to a cosmological constant $\Lambda$). To quote \cite{wolf}(2023)
\begin{quote}
\begin{it}
There is compelling evidence that the Universe is undergoing a late phase of accelerated expansion.
One of the simplest explanations for this behaviour is the presence of dark energy. A plethora of
microphysical models for dark energy have been proposed. The hope is that, with the ever increasing
precision of cosmological surveys, it will be possible to precisely pin down the model. We show that
this is unlikely and that, at best, we will have a phenomenological description for the microphysics of
dark energy. Furthermore, we argue that the current phenomenological prescriptions are ill-equipped
for shedding light on the fundamental theory of dark energy.
\end{it}
\end{quote}

As described in Section 1.1, there are two major epochs in the evolution of the antigravity universe which have quite different matter/antimatter fractions; before (BH) and after (AH) hadronization. Recall that, in the standard cosmology, hadronization occurs around a microsecond after the Big Bang. This, of course, may occur at a different time in the evolution of the antigravity universe.
\begin{itemize}

\item[~] \underline{\bf Before Hadronization}

Assuming the fundamental fermions, antifermions, and intermediate vector bosons were in some kind of thermal equilibrium just after the Big Bang, then right from beginning the matter-antimatter ratio was on the order of 2:1. This would be the period of maximum accelerated expansion. If one associates the antifermion mass-energy with the cosmological constant, then this would argue that in the BH epoch $\rho_\Lambda\sim\rho_m/2$. The Hubble parameter would also be affected by this distribution of matter and antimatter. Almost certainly the Robertson-Walker metric (eq. 3) is not the complete solution if antigravity is active and so equations (5) and (6) would also have to be modified in some way. Hence it's difficult to say without having these modified equations exactly what will quantitatively be the affect on cosmic expansion.

\item[~] \underline{\bf After Hadronization}

Now almost all of the antimatter due to antiquarks has been subsumed into antinucleons which are more matter than antimatter. From a gravitational standpoint, this effectively increases the "baryonic" matter density by roughly one-third and essentially eliminates the antimatter contribution of antiquarks. As well, many of the positrons are no longer "free" antimatter due to their combination with antinucleons to form antihydrogen and antihelium. From this it follows that the cosmic acceleration (i.e., the dark energy content) would be reduced from its BH value. This gets slightly complicated by the fact that once Black Holes are produced then the antineutrino content, and hence the "dark energy" content, of the universe would be continuously increasing albeit by an amount that may be negligible compared to the AH acceleration rate and so perhaps this may not be observable. The Hubble parameter would also be different from the BH value since the distribution of matter has changed.
\end{itemize}

The affect of hadronization on the relative amounts of matter and antimatter in the universe then leads to the following;
1) there is more repulsive energy (i.e., dark energy) in the early universe (the BH Epoch) than after hadronization and 2) the Hubble parameter will have different values in the BH and AH Epochs. 

One of the proposals put forward for generating a varying $H_0$ is the idea of "Early Dark Energy" (EDE), an extra dark energy source which exists before recombination (see \cite{EDE}(2023) for a review). Very recently it has been claimed that EDE can both resolve the Hubble tension and account for earlier than expected galaxy formation\cite{shen}(2024). The main point to take from this section is that, opposed to having to put in ad hoc an additional amount of Dark Energy prior to recombination, in the antigravity universe it follows naturally that there is effectively more "Dark Energy" (i.e., repulsive gravity) before {\it hadronization}, so even earlier in the evolution of the universe than at the time of {\it recombination}.
\subsubsection{Dark Matter}
       
 Dark Matter is a major component of the $\Lambda$CDM description of the universe (it's in the name!). There was initially a great deal of resistance to the idea since it required the existence of a substance (particle) which is not part of the Standard Model of particle physics and which interacted only gravitationally. The main observations which argue for the existence of Dark Matter are galactic rotation curves, the distribution of matter derived from gravitational lensing analyses, and diffusion damping during recombination. 

For completeness, I should add that there is a way, favoured by some, to explain the galactic rotation curves without the need for "hidden mass" (i.e., Dark Matter) -- Modified Newtonian Dynamics (MOND)\cite{mil}\cite{bek}(1983). The key idea in MOND is that the gravitational acceleration is "acceleration modified." That is, Milgrom\cite{mil} proposed the following rule for the "true" gravitational acceleration $\vec{g}$: 
$$\vec{g}\mu(g/a_0)=\vec{g}_N$$
where $\vec{g}_N$ is the Newtonian acceleration and $a_0$ is a free parameter. The function $\mu(x)$ is unspecified but it has the limiting values of $\mu(x)=1$ for $x>>1$ and $\mu(x)=x$ for $x<<1$. 
MOND does do a good job of modelling the galactic rotation curves (even better than Dark Matter according to its adherents) and it is still the subject of much activity (see, for example, \cite{lelli}(2016), \cite{sanders}(2018), \cite{schom}(2019), and \cite{mist}(2024)). There is now also a relativistic version, so-called rMOND\cite{rMOND}(2021). The main objection to it is that it replaces one ad hoc, theoretically unmotivated idea -- Dark Matter -- with another one -- the modification of gravity at large scales. Or, another way to put it is it replaces one arbitrary parameter -- the mass of the Dark Matter particle -- with another -- the acceleration constant $a_0$. 

As mentioned, in the antigravity universe a possible scenario is one where galaxies are immersed in a diffuse cloud of antihydrogen (i.e., this would be the so-called Dark Matter halo). Immediately one has from this that $\rho_{\rm DM}=\rho_{\bar{H}}\sim\rho_H$. That is, as observed, the Dark Matter density (which is now due to the antihydrogen cloud) would appear to be roughly equal to the baryon density. The antihydrogen/hydrogen (same as the antinucleon/nucleon) relative gravitational interactions are given in eq. (2). This comes from the fact that the gravitational mass of hydrogen, $m_G$, is equal to the inertial mass, $m_I$ ($\equiv m_H$), while the gravitational mass of antihydrogen, $\bar{m}_G$, is effectively around a third of the inertial mass of antihydrogen, $\bar{m}_I$, which, from $CPT$ Invariance, is equal to the inertial mass of hydrogen (i.e., $\bar{m}_I$=$m_I$=$m_H$). To see how this could mimic a MOND-like solution to galactic rotation curves it is necessary to rearrange eq (2). That is,

\begin{eqnarray*}
    F_{HH}&=&-G\frac{m_G^2}{r^2}~=~-G\frac{m_I^2}{r^2}\\
    &\equiv&-G_{HH}\frac{m_H^2}{r^2}\\
    F_{H\bar{H}}&=&-G\frac{m_G\bar{m}_G}{r^2}~\approx~-G\frac{m_I(\bar{m}_I/3)}{r^2}~=~-\frac{G}{3}\frac{m_I^2}{r^2}\\
    &\equiv&-G_{H\bar{H}}\frac{m_H^2}{r^2}\\
    F_{\bar{H}\bar{H}}&=&-G\frac{\bar{m}_G\bar{m}_G}{r^2}~\approx~-G\frac{(\bar{m}_I/3)(\bar{m}_I/3)}{r^2}~=~-\frac{G}{9}\frac{m_I^2}{r^2}\\
    &\equiv&-G_{\bar{H}\bar{H}}\frac{m_H^2}{r^2}
\end{eqnarray*}
Written in this way we now have distributions of hydrogen and antihydrogen atoms, all with mass $m_H$, with effective gravitational couplings of $G_{HH}=G$, $G_{H\bar{H}}\approx G/3$, and $G_{\bar{H}\bar{H}}\approx G/9$. 

What one then needs to know is the spatial distribution of the stars, dust, and gas as well as the antihydrogen cloud (which presumably extends far out past the boundary of the galactic matter). So one potentially has the following scenario for a galaxy; a tightly bound core (coupling is $G$) of hydrogen and helium in the form of stars, dust, and gas surrounded by (immersed in?) a diffuse cloud of antihydrogen which has a gravitational coupling of $\sim G/3$ with the core and $\sim G/9$ with the surrounding antihydrogen atoms. Again, the beauty of this is that it follows naturally in the antigravity universe, there is no need to add any extra particles or play with the force of gravity. 

As discussed in Section 2.2, any conclusions based on lensing analyses aren't strictly valid in the antigravity universe and would have to be redone. There has been one analysis relevant for damping\cite{sergey}(2024)\footnote{See \cite{jeans}(2014) for a nice discussion of the Jeans Instability as applied to gravity.} 
that incorporates the gravitational properties of antihydrogen introduced in \cite{antiart}. They conclude that, "In general, the detailed
analysis of GAI\footnote{In \cite{sergey} GAI is an acronym for "gravitational asymmetry index". I called it the antimatter fraction in \cite{antiart}.} for baryons in the framework of quantum chromodynamics can give the correct understanding
of the experimental observation of falling antihydrogen and in general on the existence of repulsive gravity." 


   \subsubsection{Cosmic (Anti)Voids}
   If there is a "smoking gun" for the antigravity universe it is the absolute necessity that there be cosmic voids. The vast number of antineutrinos which repel everything except each other would necessarily occupy regions of space devoid of matter. As a matter of fact, there would naturally develop bubbles of antineutrinos since the repellant force of surrounding matter (including antinucleons) is much, much greater than the attractive force between antineutrinos. This follows straightforwardly from equation (1) where one finds;
   $$F_{\bar{\nu}\bar{\nu}}=-G\frac{m_{\bar{\nu}}^2}{r^2}=-G\left(\frac{m_{\bar{\nu}}}{m_N}\right)\frac{m_{\bar{\nu}}m_N}{r^2}=-\left(\frac{m_{\bar{\nu}}}{m_N}\right)F_{\bar{\nu}N}=-3\left(\frac{m_{\bar{\nu}}}{m_N}\right)F_{\bar{\nu}\bar{N}}$$
   which, with $m_N\sim 1$ GeV and $m_{\bar{\nu}}\sim 1$ meV, means that $F_{\bar{\nu}N} \sim -10^{12}F_{\bar{\nu}\bar{\nu}}$. The antineutrinos would therefore act like a weakly-interacting, confined gas trapped within matter "walls". These bubbles could also be regions of trapped positrons since, as above, 
   $$F_{e^+N}=3F_{e^+\bar{N}}=-\left(\frac{m_N}{m_e}\right)F_{e^+e^+}\approx -10^3F_{e^+e^+}$$
   However there are far fewer positrons than antineutrinos since many of the them would be absorbed into antihydrogen plus some of the free positrons would presumably "leak out" of the void as the "wall" pressure is much less than that for antineutrinos. They also are electrically repelling each other. 

   Having said all that, these are not the cosmic voids 
   discussed in \cite{cont1}(2024)
   which are basically regions of space containing very little in the way of anything\footnote{As a matter of fact, cosmological voids seem to be more empty of galaxies than
expected, as pointed out by Peebles\cite{peebles}(2001).}. 
These bubbles filled with antineutrinos (which to avoid confusion with the standard term I'll call antivoids) are actually full of antineutrino mass-energy. Cosmic antivoids would observationally appear similar to voids since the antineutrinos are weakly interacting so it would seem like the space is essentially empty. However, in the antigravity universe antineutrinos gravitationally {\it deflect} photons complicating the observation of galaxies behind the antivoids. So the evidence for voids is still also evidence for antivoids but the conclusions as to their spatial extent and contributions to cosmological parameters may be different. Furthermore, if the voids are actually antivoids, then various analyses like that presented in \cite{pisano}(2024) need to be rethought as there is presumably little to no matter in the form of stars or galaxies to be found in the antivoids.
   \subsubsection{The Cosmic Microwave Background}
One of the key observational backbones of $\Lambda$CDM is the CMB and the precision measurements from {\it Planck}, either alone or combined with data from other observatories, puts tight constraints on cosmological parameters\cite{planck}(2024). One thing to keep in mind is that CMB photons are, of course, affected by lensing as they pass by gravitational potential wells. This is a source of systematic error in standard CMB analyses and was first characterized through the introduction of the so-called lens parameter (originally as $A_{lens}$ in \cite{smoot}(2008) and now $A_L$). The lens parameter for GR has the value 1 while you would get a value of 0 for no lensing. Early {\it Planck} analyses found a preference for $A_{lens}>1$ but the recent analysis\cite{planck} finds "$A_L$ = $1.039\pm 0.052$, which aligns more closely
with theoretical expectations within the $\Lambda$CDM framework." In any case, again, in the antigravity universe, it is extraction of parameters using lensing that will differ from the $\Lambda$CDM values for any set of observations. That is, a CMB analysis in the antigravity universe may result in a somewhat different value for, for example, the "clumpiness" parameter $S_8$ (or its equivalent) but the the overall message from the CMB will not be radically different.  
   
   \subsubsection{X-rays from Clusters} 
   Characteristic x-rays from galaxy clusters (GCs) is an observable used in $\Lambda$CDM analyses to extract cosmological parameters (see, for example, \cite{wan}(2021) or \cite{ghir}(2024)). The general principles underlying the x-ray production have been understood for some time. As stated in \cite{serm}(1977), "The data are adequately described by emission from an isothermal plasma with an iron abundance
in near agreement with cosmic levels." The x-rays are produced in the intracluster medium (ICM), a hot plasma composed of supernovae remnants (star "dust") gravitationally compressed by the surrounding galaxies and Dark Matter\footnote{Although galaxy clusters are understood in general, there is research being done as to the specific morphologies, as discussed in \cite{kafer} (2019), for example.}. The situation is somewhat, though not radically, different in the antigravity universe. There would certainly be GCs but, since there are little to no anti-stars, there are no anti-galaxy clusters and the ICM of a GC does not contain any anti-supernova remnants (antidust). So the x-rays observed are indeed evidence of GCs and are presumably produced in the recognized fashion. What does complicate matters is that the ICM most certainly would contain antihydrogen, possibly all ionized due to the assumed high temperature of the ICM ($\sim$$10^7~^\circ$K). I say assumed because like with all $\Lambda$CDM calculations there are various assumptions made about the GC. A key ingredient needed for extracting cosmological parameters is the mass of the GC and this is generally derived from lensing data, as was done in \cite{ghir}. And a recurring theme of this paper is that lensing data needs to be reanalyzed in the antigravity universe to account for the large amounts of antihydrogen and antineutrinos (and, perhaps, also positrons) present.

   \subsubsection{Lyman-$\alpha$ Forest}
   The so-called "Lyman-$\alpha$ Forest" (Ly$\alpha$F), the red-shifted distribution of hydrogen absorption lines, is another observable which is widely used to extract $\Lambda$CDM parameters as well as constrain proposed solutions to discrepencies in $\Lambda$CDM (see, for example, \cite{gold}(2023)). Such analyses will be clearly impacted by the premise that galaxies (composed of stars, gas, and dust) are immersed in a diffuse cloud of antihydrogen atoms. There are a number of conditions underlying the Ly$\alpha$F which are characteristic of galaxies (or quasistellar objects, QSO) but not to the antihydrogen cloud, the lack of antistars and antidust being the major one, so the exact amount of "line-of-sight" hydrogen+antihydrogen has a large uncertainty. It is true that the spectral lines of antihydrogen are expected to be identical to those of hydrogen\footnote{We showed in the ALPHA experiment that the energy levels of antihydrogen are equivalent to those of hydrogen to 12 decimal places\cite{ALPHA}.}. There's also the question of photoionization of the antihydrogen due to the cosmic UV background. Perhaps the spatial distribution of the antihydrogen required to mimic Dark Matter (if it exists) is incompatible with Ly$\alpha$F observations. Again (unsatisfyingly), calculating the exact effect requires assumptions about the spatial distribution of the antihydrogen. 

   As an aside, not only is Lyman-$\alpha$ absorption used to study galaxies but also Lyman $\alpha$ emission gives information on galaxy formation (see, for example, \cite{cant}(2012)). A recent study\cite{lamb} observed a lack of Lyman $\alpha$ emitters around a quasar leading to speculation on ideas of star formation around supermassive black holes in the early universe. The galactic halos of antihydrogen in the antigravity universe would certainly require a reinterpretation of Lyman $\alpha$ measurements.     


      \subsubsection{Long-lasting Gamma-Ray Bursts from Black Holes}
The influence of a highly-negatively-charged Black Hole (discussed in Appendix A) on surrounding nucleons and antinucleons is more subtle than one would naively surmise. Consider first the proton and antiproton. The proton is positively charged and completely composed of matter so it would be both electrically and gravitationally attracted to the negatively-charged, matter-dominated Black Hole. The antiproton is two-thirds matter and one-third antimatter and is negativly charged. Hence it is gravitationally attracted to a Black Hole but electrically repelled by it. So while the two forces are not equal it seems reasonable to assume that the net force on the antiproton will be much, much weaker than that for the proton and may be attractive or repulsive. The neutron, which is fully matter, and the antineutron, roughly two-thirds matter and one-third antimatter, are both attracted to the Black Hole\footnote{In \cite{bambi} they describe the scenario where very low mass Black Holes ($\sim 10^{20}$ gm) become positively charged and, hence, are prodigious sources of positrons.}.  

The fact that a Black Hole is highly negatively charged means that the hydrogen and helium from a nearby star would be quickly ionized and the remaining protons and alpha particles would become powerful synchrotron light emitters as they accelerate towards the Black Hole. A proper calculation needs to be done but this could potentially be the source of long-lasting gamma-ray bursts from Black Holes\cite{ho}(2023). In contrast, as explained above, antiprotons and antialpha particles should emit little or no synchrotron light since the net force on them is considerably smaller. 
\subsection{Self-Consistency}
The power of the $\Lambda$CDM, at least at the turn of the century, was that it integrated a plethora of observations into a coherent, consistent picture of the universe. That is, the correlation of a number of observables has been used to put powerful constraints on cosmological parameters. For example, baryon acoustic oscillations (BAO) in the Ly$\alpha$F fluctuations\cite{BOSS}(2013)\cite{BLy}(2019) combined with CMB lensing data from {\it Planck} and ACT was used to extract values for $H_0$ and $\Omega_m$\cite{DESI}(2024). Or there is an analysis combining DESI measurements with CMB lensing results from {\it Planck} and ACT to extract values of $S_8$ and $\sigma_8$\cite{DPA}(2024). There are many analyses like this combining different measurements and generally finding consistent parameter sets thus lending power to the $\Lambda$CDM model. But there are still some niggling issues with $\Lambda$CDM, like a varying Hubble parameter or an expansion acceleration changing over time, as well as fundamental ones, like the nature of Dark Matter, which I have tried to give a flavour of in the previous Sections. The parameters to be fit for in the antigravity universe may very well be different than those of $\Lambda$CDM but the overarching question I'm posing in this article is, "Would the antigravity universe also allow for the various observations to be integrated into a consistent picture?" I've not provided an answer to that question, other than in a hand-wavy way\footnote{You're no doubt tired of reading the phrase, "A proper calculation needs to be done ..."}, and so the need for an ...

\subsection{Apologia}
I wrote this article in the spirit of the 1898(!) letter to {\it Nature}\cite{dream} by Arthur Schuster in which he proposed the idea of what he termed "antimatter", a substance which is repelled gravitationally by matter. He concluded his paper with the words, "Whether
such thoughts are ridiculed as the inspirations of madness, or allowed to be the serious possibilities of a future science, they add renewed interest to the careful examination of the incipient worlds which our telescopes have revealed to us. Astronomy, the oldest and yet most juvenile of sciences, may still have some surprises in store. May antimatter be commended to its care!" 

Even given that, I'm fully cognizant of the fact that I have seriously failed (at least) one of R. Ehrlich's criteria for an idea to be taken seriously\cite{Ehrlich}. Specifically I'm referring to number 8 -- "{\it The idea should not try to explain too much or too little.}" And I'm probably also on pretty thin ice with regard to number 2 -- "{\it The proposer of the idea should be a knowledgeable and respected scientist, although he or she may come from outside that particular field.}" But I believe I'm on firmer ground with respect to a number of the other criteria including; 
\begin{itemize}
    \item[1.] {\it The new hypothesis should make some contact with familiar physics.}\\
    The physics discussed here is that of GR and Newtonian gravity.
    \item[5.] {\it The proposer should have no agenda going beyond the science of the issue.}\\
    I can only state that this is true.
    \item[6.] {\it The theory should not have many free parameters.}\\
    As it stands the only free parameter is $G$. 
\end{itemize}

Perhaps the strongest argument for the version of the antigravity universe presented here is that it does satisfy criterion 10 -- "{\it The theory should be the simplest explanation of the phenomena.}" It's easy to forget, given its present status as the accepted description of the universe, just how profoundly weird the $\Lambda$CDM picture is. It was not driven by underlying theoretical ideas but purely by observation. This is of course how science should work but $\Lambda$CDM does require the introduction of a wholly new substance, Dark Matter, with properties different from any Standard Model particle, as well as Dark Energy, which comes with its own difficulties (the "Naturalness Problem"). There's also the need for some method, as yet undiscovered, for generating the baryon-antibaryon asymmetry. In principle, although again maybe not in practice, the antigravity universe as outlined requires none of this additional physics. The postulates upon which the antigravity universe is built are; 1) A version, or extension, of GR exists in which matter and antimatter repel\footnote{The simplest and most elegant version of this is due to Villata\cite{vill1} where antigravity is inherently a part of GR if $CPT$ is conserved.} and 2) the mass-energy of the fundamental antifermions (antiquarks and antileptons) act gravitationally as antimatter while the mass-energy of the fundamental fermions (quarks and leptons) and the fundamental intermediate vector bosons (e.g., photon, gluon, etc.) act gravitationally as matter. That's it. 

Finally, there is one of the additional criteria suggested by Sanders\cite{sand}(2015), "{\it First of all, it is significant if the idea is falsifiable.}" The antigravity universe is certainly falsifiable and will most probably be falsified within the next year when we (ALPHA-g) publish an updated measurement of the free-fall acceleration of antihydrogen and find that it is compatible with expectations (i.e., $a_{\bar{H}}=g$) and incompatible with the prediction from \cite{antiart} of $a_{\bar{H}}=(0.33^{+0.23}_{-0.11})g$. I readily concede that there is a paucity of calculations included in this paper but it seems to me premature (foolhardy?) to attempt any type of serious calculation connecting observables to antigravity universe parameters until after the newest ALPHA-g result comes out. This situation will undoubtedly change if $a_{\bar{H}}$ was found to be significantly different from $g$. 

\section{Conclusions}

In this article I've tried to flesh out a picture of the universe in which antigravity is acting and matter and antimatter are precisely defined. 
Did this allow for an answer to the question posed in the title of this article? The best that can be said at present without a fully predictive version of antigravity is "maybe." But it has been fun to speculate about it.
\appendix

\section{Black Holes -- Highly Charged Antineutrino Factories}
  Another object which leads to unique consequences in the antigravity universe is a Black Hole. For example, Black Holes would become negatively charged since the positively-charged antileptons (e.g., the positron) produced in Hawking radiation would be repelled by a Black Hole. This is true whether the Black Hole is wholely matter if comprised of nucleons or dominantly matter if built from antinucleons (or some combination of the two). So Black Holes should become more and more negatively charged with time until presumably they become negatively charged enough that they also electrically repel negatively-charged leptons (e.g., $\mu^-$) and electrically attract positively-charged antileptons. I suppose some sort of equilibrium state is quickly reached. For charged lepton-antilepton pair equilibrium at the event horizon we'd need something like:
   $$GM_{\rm BH}m_\ell\approx\frac{1}{4\pi\varepsilon_0}Q_{\rm BH}e$$
where I assumed the Black Hole (like a cow) is a sphere. Now, taking $Q_{\rm BH}=N_ee$ and $M_{\rm BH}=BM_\odot$ then, for example, the stability equation for electron-positron pairs is:
$$\frac{N_e}{B}=\frac{GM_\odot m_e}{(1/4\pi\varepsilon_0)e^2}
=5.2\times 10^{35}$$
   That is, the charge of the Black Hole would be $Q_{\rm BH}\approx -(5\times 10^{35})Be$. Note that if there was some mechanism which reduced this charge (e.g., due to the production of other charged pairs or ionization of surrounding matter) then a Black Hole would be a source of positrons, the observation of which is described in Section 1.2.2.

   No equilibrium between electrical and gravitational forces would be achieved in the case of the electrically neutral neutrinos and so Black Holes would be prodigious sources of antineutrinos produced via Hawking radiation as they would be repelled by the matter-dominated Black Hole.  

\section{Antistar Annihilation}

For the purposes of discussion of particle-antiparticle annihilation, the antiproton is not the antiparticle of the proton. It is the quarks and antiquarks that annihilate so one not only expects the antiproton ($\bar{u}\bar{u}\bar{d}$) to "annihilate" with the proton ($uud$) but also with the neutron ($udd$) (and, likewise, an antineutron with both a proton and a neutron). Naively, from counting quark-antiquark pairs, one would expect the antiproton-proton annihilation rate to be something like 1.25 times the antiproton-neutron rate. That is, if you label identical quarks, we have the
 $\bar{p}$ = ($\bar{u}_1$$\bar{u}_2$$\bar{d}$), the
 $p$ = ($u_1u_2d$), and the
 $n$ = ($ud_1d_2$).
So there are 5 different $\bar{p}p$ $\bar{q}q$ annihilation channels --
$\bar{u}_1u_1$, $\bar{u}_1u_2$, $\bar{u}_2u_1$, $\bar{u}_2u_2$, and $\bar{d}d$ --
whereas there are 4 different $\bar{p}n$ $\bar{q}q$ annihilation channels --
$\bar{d}d_1$, $\bar{d}d_2$, $\bar{u}_1u$ and $\bar{u}_2u$. Hence you have $N(\bar{p}p)/N(\bar{p}n)$ = 5/4 = 1.25. 
 This naive estimate turns out to describe what happens quite well. In a 1966 paper\cite{chin} looking at antiprotons annihilating at rest on deuterium, "the ratio of the number of annihilations [of antiprotons] on protons to that on neutrons was found to be $1.3\pm 0.07$." 

Antinucleon-nucleon annihilation results in the production of pions\footnote{Some $\sim$ 6\% of annihilations produce kaons but this doesn't change the picture presented here in any significant way.}. The $\pi^\pm$ and $\pi^0$ masses\footnote{All branching fractions, masses, and lifetimes are taken from the PDG summary tables\cite{pdg}.} are $139.57039 \pm 0.00018$ MeV and $134.9768\pm 0.0005$ MeV, respectively.
From \cite{klempt} we have that the mean number of pions per annihilation is:
$$n_\pi = 4.98 \pm 0.13~~ n_{\pi^\pm} = 3.05 \pm 0.04~~ n_{\pi^0} = 1.93 \pm 0.12$$
The subsequent dominant pion decay channels are:
$$N\bar{N}\longrightarrow\cases{\pi^0\longrightarrow\gamma\gamma~~&(98.8\%)\cr 
\pi^+\longrightarrow\mu^+\nu_\mu~~&(99.99\%)\cr \pi^-\longrightarrow\mu^-\bar{\nu}_\mu~~&(99.99\%)}$$
This leads to two distinct types of evidence for antinucleon-nucleon annihilation:
\subsection{Charged Pions}
As described, the charged pions decay 99.99\% of the time to a muon and a muon neutrino
with a lifetime of 26 ns. The muons then decay via $\mu^+\rightarrow e^+\bar{\nu}_\mu\nu_e$ and  $\mu^-\rightarrow e^-\nu_\mu\bar{\nu}_e$ with a lifetime of 2.2 $\mu$s. The main point is that each annihilation leads to the production of a number of neutrinos and, more importantly for the antigravity universe, antineutrinos.
\subsection{Neutral Pions}
The $\pi^0$ decays essentially instantly ($\tau=8\times 10^{-17}$ s) to two photons 
with each photon therefore having a $\pi^0$ rest-frame energy of $m_{\pi^0}/2$ $\sim 67.5$ MeV. The maximum pion momentum from $N\bar{N}$ annihilation at rest is $\sim$930 MeV (for the $\pi\pi$ final state) while the average pion momentum is around 300 MeV. The mean number of photons per annihilation is $n_\gamma = 3.93 \pm 0.24$ (again from \cite{klempt}). The inclusive photon energy distributions (both simulated and data) from proton-antiproton annihilation are 
given in \cite{backen}. They are fairly broad, peak around 70 MeV, and end sharply at $\sim 950$ MeV. Spectra put forth as evidence for antistars in our galaxy that are shown in \cite{antistar} are taken from \cite{aspect}.

\end{document}